\documentclass[preprint,aps,showpacs,nofootinbib,preprintnumbers,amsmath,amssymb]{revtex4-1}
\usepackage{}
\usepackage{epsfig}
\usepackage{subfigure}
\usepackage{dcolumn}% Align table columns on decimal point
\usepackage{bm}% bold math
\usepackage[usenames ,dvipsnames]{xcolor}
\usepackage{slashed}
\usepackage{graphicx,color}

\begin{document}
\title{Charmful two-body anti-triplet $b$-baryon decays}

\author{Y.K. Hsiao$^{1,2,3}$, P.Y. Lin$^{4}$, C.C. Lih$^{5,2,3}$, and C.Q. Geng$^{1,2,3}$}
\affiliation{
$^{1}$Chongqing University of Posts \& Telecommunications, Chongqing, 400065, China\\
$^{2}$Physics Division, National Center for Theoretical Sciences, Hsinchu, Taiwan 300\\
$^{3}$Department of Physics, National Tsing Hua University, Hsinchu, Taiwan 300\\
$^{4}$Department of Electrophysics National Chiao Tung University,   Hsinchu, Taiwan 300\\
$^{5}$Department of Optometry, Shu-Zen College of Medicine and Management, Kaohsiung Hsien,Taiwan 452}
\date{\today}

\begin{abstract}
We study the charmful decays of the two-body ${\cal B}_b\to {\cal B}_n M_c$ decays, where
${\cal B}_b$ represents the anti-triplet of $(\Lambda_b,\Xi_b^0,\Xi_b^-)$,  ${\cal B}_n$ stands for 
the baryon octet and $M_c$ denotes as the charmed meson of $D^{(*)}_{(s)}$, $\eta_c$ and $J/\psi$.
Explicitly, we predict that ${\cal B}(\Lambda_b\to D_s^- p)=(1.8\pm 0.3)\times 10^{-5}$, which is
within the measured upper bound of 
${\cal B}(\Lambda_b\to D_s^- p)<4.8(5.3)\times 10^{-4}$ at 90\%\,(95\%) C.L., and reproduce
${\cal B}(\Lambda_b\to J/\psi \Lambda)=(3.3\pm 2.0)\times 10^{-4}$ and 
${\cal B}(\Xi_b^-\to J/\psi \Xi^-)=(5.1\pm 3.2)\times 10^{-4}$ in agreement with the data.
Moreover, we find that
${\cal B}(\Lambda_b \to \Lambda \eta_c)=(1.5 \pm 0.9) \times 10^{-4}$,
${\cal B}(\Xi_b^- \to \Xi^- \eta_c)=(2.4 \pm 1.5) \times 10^{-4}$ and 
${\cal B}(\Xi_b^0 \to \Xi^0 \eta_c,\Xi^0 J/\psi)=(2.3 \pm 1.4,4.9 \pm 3.0) \times 10^{-4}$,
which are accessible to the experiments at the LHCb.
\end{abstract}
%\pacs{}

\maketitle
\section{introduction}
The two-body decays of
$\Lambda_b\to \Lambda_c^+ K^-$, $\Lambda_c^+ \pi^-$, 
$\Lambda_c^+ D^-$, and $\Lambda_c^+ D_s^-$
can be viewed as through the $\Lambda_b\to \Lambda_c$ transition along with
the recoiled mesons $K^-$, $\pi^-$, $D^-_s$ and $D^-$, respectively, such that 
one may use the factorization ansatz to get the fractions of the branching ratios as
\begin{eqnarray}
{\cal R}_{K/\pi}&\equiv&
\frac{{\cal B}(\Lambda_b\to \Lambda_c^+ K^-)}{{\cal B}(\Lambda_b\to \Lambda_c^+\pi^-)}
\simeq \frac{(|V_{us}|f_K)^2}{(|V_{ud}|f_\pi)^2}=0.073\,,\nonumber\\
{\cal R}_{D/D_s}&\equiv&
\frac{{\cal B}(\Lambda_b\to \Lambda_c^+ D^-)}{{\cal B}(\Lambda_b\to \Lambda_c^+D_s^-)}
\simeq \frac{(|V_{cd}|f_D)^2}{(|V_{cs}|f_{D_s})^2}=0.034\,,
\end{eqnarray}
which are in agreement with the data, given by~\cite{Aaij:2013pka,Aaij:2014pha}
\begin{eqnarray}\label{RKpi}
{\cal R}_{K/\pi}=0.0731\pm 0.0016\pm 0.0016\,,\;
{\cal R}_{D/D_s}=0.042\pm 0.003\pm 0.003\,.
\end{eqnarray}
In the same picture, the measured ${\cal B}(\Lambda_b\to pK^-,p\pi)$
can be also explained~\cite{pdg,Hsiao:2014mua}.
In addition, the direct CP violating asymmetry of $\Lambda_b\to p K^{*-}$ is predicted
as large as 20\%~\cite{Geng:2006jt}.  

On the other hand, the branching ratios of $\Lambda_b\to D^-_s p$, 
$\Lambda_b\to J/\psi \Lambda$ and $\Xi_b^-\to J/\psi \Xi^-$
are shown as~\cite{pdg,Aaij:2014lpa}
\begin{eqnarray}
\label{Bb_fBb}
{\cal B}(\Lambda_b\to D^-_s p)&=&
(2.7\pm 1.4\pm 0.2\pm 0.7\pm 0.1\pm 0.1)\times 10^{-4}\; \text{or} \nonumber\\
&<&4.8(5.3)\times 10^{-4}\;\text{at 90\%\,(95\%) C.L.}\,,\nonumber\\
{\cal B}(\Lambda_b\to J/\psi \Lambda)&=&(3.0\pm 1.1)\times 10^{-4}\,,\nonumber\\
{\cal B}(\Xi_b^-\to J/\psi \Xi^-)&=&(2.0\pm 0.9)\times 10^{-4}\,,
\end{eqnarray}
with ${\cal B}(\Lambda_b\to J/\psi \Lambda)$ and ${\cal B}(\Xi_b^-\to J/\psi \Xi^-)$
converted from the partial observations of
${\cal B}(\Lambda_b\to J/\psi \Lambda)f_{\Lambda_b}=(5.8\pm 0.8)\times 10^{-5}$ and
${\cal B}(\Xi_b^-\to J/\psi \Xi^-)f_{\Xi_b^-}=(1.02^{+0.26}_{-0.21})\times 10^{-5}$, where
$f_{\Lambda_b}=0.175\pm 0.106$ and $f_{\Xi_b}=0.019\pm 0.013$ are the fragmentation fractions of 
the $b$ quark to $b$-baryons of $\Lambda_b$ and $\Xi_b$~\cite{fLb}, respectively. 
Nonetheless, for these  ${\cal B}_b\to {\cal B}_n M_c$ decays in Eq.~(\ref{Bb_fBb}),
%where ${\cal B}_n$ and $M_c$ stand for
%the baryon octet and the charmed meson, respectively,
the theoretical understanding is still lacking.
Since the factorization approach is expected  to be reliable in studying the branching ratios of
${\cal B}_b\to {\cal B}_n M_c$, in this report, 
we shall systematically analyze the branching ratios 
for all possible ${\cal B}_b\to {\cal B}_n M_c$ decays, and 
compare them with the experimental data at the $B$-factories, as well as the LHCb,
where ${\cal B}_b$, ${\cal B}_n$ and $M_c$ correspond to 
the anti-triplet $b$-baryon of $(\Lambda_b,\Xi_b^0,\Xi_b^-)$,   
 baryon octet and  charmed meson, respectively.

\section{Formalism}
%=======================
\begin{figure}[t!]
\centering
\includegraphics[width=2.5in]{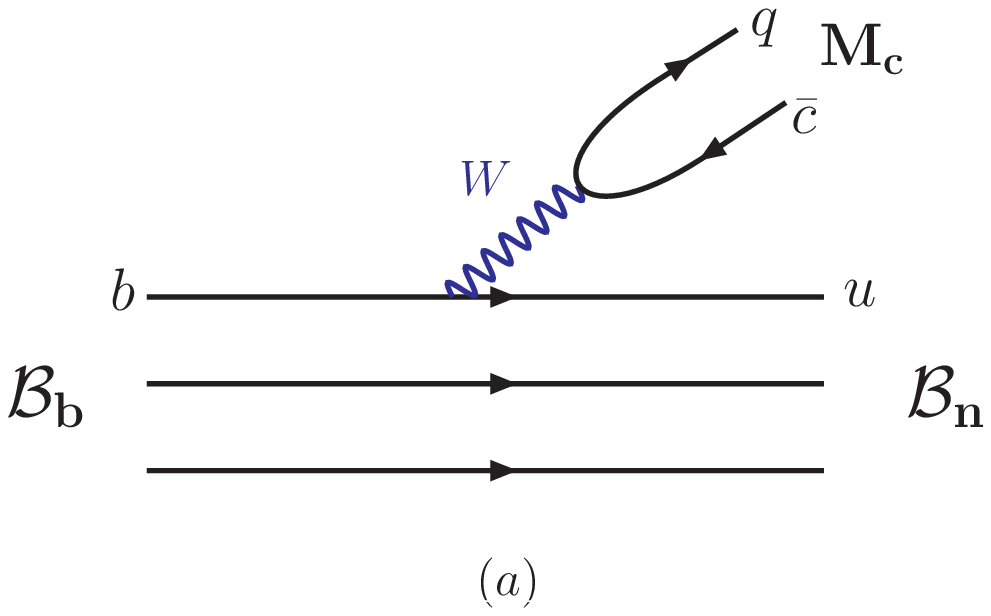}
\includegraphics[width=2.5in]{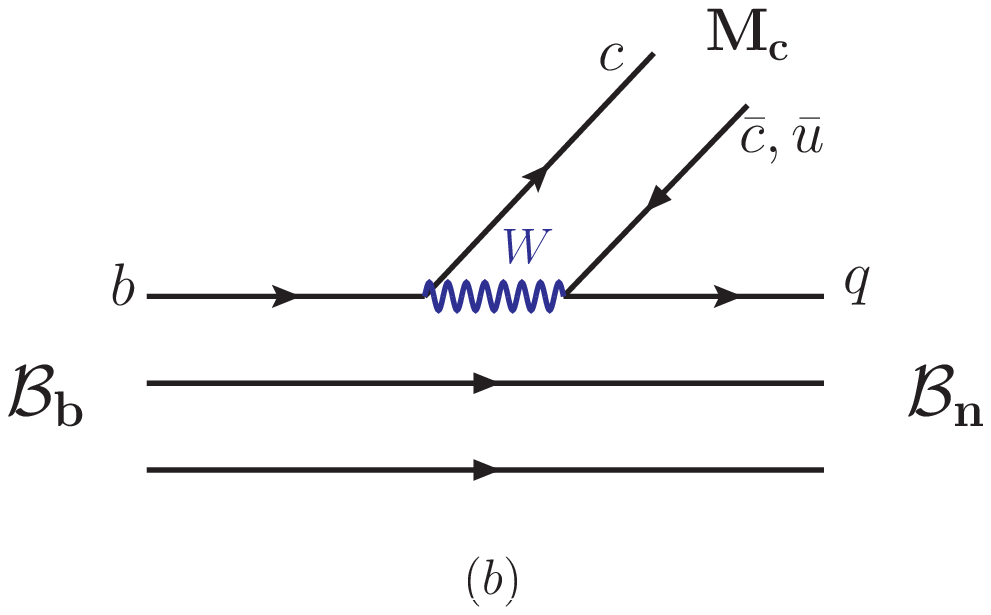}
\caption{Diagrams for two-body charmful ${\cal B}_b\to {\cal B}_n M_c$ decays.}\label{BbtoBnMc}
\end{figure}
%======================
As the studies 
in Refs.~\cite{Gutsche:2013oea,Liu:2015qfa,Wei:2009np,
Fayyazuddin:1998ap,Chou:2001bn,Ivanov:1997ra,Cheng:1996cs},
based on the factorization approach,
the amplitudes for the two-body charmful $b$-baryon decays are presented in terms of
the decaying process of the ${\cal B}_b\to {\cal B}_n$ transition with the recoiled charmed meson $M_c$. 
From Fig.~\ref{BbtoBnMc}a, the amplitudes of ${\cal B}_b\to {\cal B}_n M_c$ 
via the quark-level $b\to u\bar c q$ transition are factorized as
\begin{eqnarray}\label{eq1}
{\cal A}_1({\cal B}_b\to {\cal B}_n M_c)&=&
\frac{G_F}{\sqrt 2}V_{ub}V_{cq}^*a_1\,
\langle M_c|\bar q\gamma^\mu(1- \gamma_5) c|0\rangle
\langle {\cal B}_n|\bar u\gamma_\mu(1-\gamma_5) b|{\cal B}_b\rangle\,, 
\end{eqnarray}
where $G_F$ is the Fermi constant, $V_{ub,cq}$ are the CKM matrix elements, while
the explicit decay modes are 
\begin{eqnarray}
\Lambda_b\to p M_c,\;\Xi_b^-\to \Lambda(\Sigma^0) M_c,\;\Xi_b^0\to \Sigma^+ M_c\,
\end{eqnarray}
with $q=d(s)$ for $M_c=D^{(*)-}(D^{(*)-}_s)$. On the other hand,
the amplitudes via the quark-level $b\to c\bar u q$ ($b\to c\bar c q$) transition 
in Fig.~\ref{BbtoBnMc}b can be written as
\begin{eqnarray}\label{eq2}
{\cal A}_2({\cal B}_b\to {\cal B}_n M_c)&=&
\frac{G_F}{\sqrt 2}V_{cb}V_{q_1 q}^*a_2\,
\langle M_c|\bar c\gamma^\mu(1- \gamma_5) q_1|0\rangle
\langle {\cal B}_n|\bar q\gamma_\mu(1-\gamma_5) b|{\cal B}_b\rangle\,, 
\end{eqnarray}
with $q_1=u$ for $M_c=D^{(*)0}$ and $q_1=c$ for $M_c=\eta_c$ and $J/\psi$,
where the decays of ${\cal B}_b\to {\cal B}_n M_c$ are
\begin{eqnarray}
&&\Lambda_b\to n M_c,\;
\Xi_b^-\to \Sigma^-M_c,\;\Xi_b^0\to\Lambda(\Sigma^0)M_c\;\;\text{ for $q_2=d$},\nonumber\\
&&\Lambda_b\to \Lambda(\Sigma^0) M_c,\;
\Xi_b^-\to \Xi^-M_c,\;\Xi_b^0\to\Xi^0M_c\;\;\text{ for $q_2=s$}.
\end{eqnarray}
In this study, we will exclude the study of $\Lambda_b\to n M_c$  
due to the elusive neutron in the $B$-factories.
The amplitudes ${\cal A}_{1,2}$ via the $W$-boson exchange diagrams
are led to be the color-allowed and color-suppressed processes.
The parameters $a_1$ and $a_2$ in Eqs.~(\ref{eq1}) and (\ref{eq2})
are presented as~\cite{Hamiltonian,ali}
\begin{eqnarray}\label{a12}
a_1=c_1^{eff}+\frac{c_2^{eff}}{N_c}\,,a_2=c_2^{eff}+\frac{c_1^{eff}}{N_c}\,,
\end{eqnarray}
with the effective Wilson coefficients $(c^{eff}_1,\,c^{eff}_2)=(1.168,-0.365)$, respectively,
where the color number $N_c$ 
should be taken as a floating number from $2\to\infty$
to account for the non-factorizable effects in the generalized factorization
instead of $N_c=3$.
The matrix elements for $P_c=(\eta_c,\,D)$ and $V_c=J(/\psi,\,D^{*}$) productions read
\begin{eqnarray}
\langle P_c|A^c_\mu|0\rangle&=&-if_{P_c} q_\mu\,,\nonumber\\
\langle V_c|V^c_\mu|0\rangle&=&m_{V_c} f_{V_c}\varepsilon_\mu^*\,,
\end{eqnarray}
with $V^c_\mu(A^c_\mu)=\bar q\gamma^\mu(\gamma_5) c$ or 
$\bar c\gamma^\mu(\gamma_5) q_1$, where $q_\mu$ and $\varepsilon_\mu^*$
are the four-momentum and polarization, respectively.
Those of the ${\cal B}_b\to {\cal B}_n$ baryon
transition in Eq. (\ref{eq1}) have the general forms:
\begin{eqnarray}\label{FFs}
&&\langle {\cal B}_n|\bar q \gamma_\mu b|{\cal B}_b\rangle=
\bar u_{{\cal B}_n}\bigg[f_1\gamma_\mu+\frac{f_2}{m_{{\cal B}_b}}i\sigma_{\mu\nu}q^\nu+
\frac{f_3}{m_{{\cal B}_b}}q_\mu\bigg] u_{{\cal B}_b}\,,\nonumber\\
&&\langle {\cal B}_n|\bar q \gamma_\mu\gamma_5 b|{\cal B}_b\rangle=
\bar u_{{\cal B}_n}\bigg[g_1\gamma_\mu+\frac{g_2}{m_{{\cal B}_b}}i\sigma_{\mu\nu}q^\nu+
\frac{g_3}{m_{{\cal B}_b}}q_\mu\bigg]\gamma_5 u_{{\cal B}_b}\,,
\end{eqnarray}
where $f_j$ ($g_j$) ($j=1,2,3$)  are the form factors.
We are able to relate the different ${\cal B}_b\to {\cal B}_n$ transition form factors
based on the $SU(3)$ flavor and $SU(2)$ spin symmetries,
which have been used to connect the space-like ${\cal B}_n\to {\cal B}'_n$ transition form factors
in the neutron decays~\cite{Brodsky1}, and the time-like $0\to {\cal B}_n \bar {\cal B}'_n$ baryonic form factors 
as well as the $B\to {\cal B}_n \bar {\cal B}'_n$ transition form factors
in the baryonic $B$ decays~\cite{Chua:2002wn,Chua:2002yd,Chen:2008sw,Geng:2011pw,Hsiao:2014zza}.
Specifically, 
$V^q_\mu=\bar q\gamma_\mu b$ and $A^q_\mu=\bar q\gamma_\mu\gamma_5 b$
as the two currents in Eq.~(\ref{FFs})
can be combined as the right-handed chiral current, that is, 
$J^q_{\mu,R}=(V^q_\mu+A^q_\mu)/2$. Consequently, 
we have~\cite{Brodsky1}:
\begin{eqnarray}\label{Gff1}
\langle {\cal B}_{n,\uparrow+\downarrow}|J^q_{\mu,R}|{\cal B}_{b,\uparrow+\downarrow}\rangle=
\bar u_{{\cal B}_n}\bigg[\gamma_\mu \frac{1+\gamma_5}{2}G^\uparrow(q^2)+
\gamma_\mu \frac{1-\gamma_5}{2}G^\downarrow(q^2)\bigg]u_{{\cal B}_b}\;,
\end{eqnarray}
where the baryon helicity states 
$|{\cal B}_{n(b),\uparrow+\downarrow}\rangle\equiv 
|{\cal B}_{n(b),\uparrow}\rangle+|{\cal B}_{n(b),\downarrow}\rangle$ 
are regarded as 
the baryon chiral states $|{\cal B}_{n(b),R+L}\rangle$ at the large momentum transfer,
while $G^\uparrow(q^2)$ and $G^\downarrow(q^2)$ 
are the right-handed and left-handed form factors, defined by
\begin{eqnarray}\label{Gff2}
G^\uparrow(q^2)=e^\uparrow_{||}G_{||}(q^2)+e^\uparrow_{\overline{||}}G_{\overline{||}}(q^2)\;,\;\;
G^\downarrow(q^2)=e^\downarrow_{||}G_{||}(q^2)+e^\downarrow_{\overline{||}}G_{\overline{||}}(q^2)\;,
\end{eqnarray}
with the constants $e^{\uparrow}_{||(\overline{||})}$ and $e^{\downarrow}_{||(\overline{||})}$
to sum over the chiral charges via the ${\cal B}_b\to {\cal B}_n$ transition, given by
\begin{eqnarray}\label{Gff3}
e^{\uparrow}_{||}             
=\langle {\cal B}_{n,\uparrow}|{\bf Q_{||}}|{\cal B}_{b,\uparrow}\rangle\;,\;\;
e^{\uparrow}_{\overline{||}}  
=\langle {\cal B}_{n,\uparrow}|{\bf Q_{\overline{||}}}|{\cal B}_{b,\uparrow}\rangle\;,\nonumber\\
e^{\downarrow}_{||}           
=\langle {\cal B}_{n,\downarrow}|{\bf Q_{||}}|{\cal B}_{b,\downarrow}\rangle\;,\;\;
e^{\downarrow}_{\overline{||}}
=\langle {\cal B}_{n,\downarrow}|{\bf Q_{\overline{||}}}|{\cal B}_{b,\downarrow}\rangle\;.
\end{eqnarray}
Note that ${\bf Q_{||(\overline{||})}}=\sum_i Q_{||(\overline{||})}(i)$
with $i=1,2,3$ as the the chiral charge operators are from
$Q_{R}^q\equiv J^q_{0,R}=q_R^\dagger b_R$,
 converting the $b$ quark in $|{\cal B}_{b,\uparrow,\downarrow}\rangle$
into the $q$ one, while the converted $q$ quark can be parallel or antiparallel to the ${\cal B}_b$'s helicity,
denoted as the subscript ($||$ or $\overline{||}$).
By comparing Eq.~(\ref{FFs}) with Eqs.~(\ref{Gff1}), (\ref{Gff2}), and (\ref{Gff3}),
we obtain
\begin{eqnarray}
f_1&=&(e^\uparrow_{||}+e^\downarrow_{||})G_{||}+(e^\uparrow_{\overline{||}}+e^\downarrow_{\overline{||}})G_{\overline{||}}\;,\nonumber\\
g_1&=&(e^\uparrow_{||}-e^\downarrow_{||})G_{||}+(e^\uparrow_{\overline{||}}-e^\downarrow_{\overline{||}})G_{\overline{||}}\;,
\end{eqnarray}
with $f_{2,3}=0$ and $g_{2,3}=0$ due to the helicity conservation,
as those derived in Refs.~\cite{CF,Wei:2009np,Gutsche:2013oea}.
It is interesting to see that, as the helicity-flip terms,
the theoretical calculations from the loop contributions
to $f_{2,3}$ ($g_{2,3}$) indeed result in the values to be
one order of magnitude smaller than that of $f_1(g_1)$,
which can be safely neglected.
In the double-pole momentum dependences, 
$f_1$ and $g_1$ can be given as~\cite{Hsiao:2014mua}
\begin{eqnarray}
f_1(q^2)=\frac{f_1(0)}{(1-q^2/m_{{\cal B}_b}^2)^2}\,,\;
g_1(q^2)=\frac{g_1(0)}{(1-q^2/m_{{\cal B}_b}^2)^2}\,,
\end{eqnarray}
such that it is reasonable to parameterize the chiral form factors to be 
$(1-q^2/m_{{\cal B}_b}^2)^2G_{||(\overline{||})}=C_{||(\overline{||})}$.
%Since we have that 
Subsequently, from
\begin{eqnarray}
&&(e^\uparrow_{||},e^\downarrow_{||},e^\uparrow_{\overline{||}},e^\downarrow_{\overline{||}})
=(-\sqrt{3/2},0,0,0)\;\; \text{for $\langle p|J^u_{\mu,R}|\Lambda_b \rangle$},\,\nonumber\\
&&(e^\uparrow_{||},e^\downarrow_{||},e^\uparrow_{\overline{||}},e^\downarrow_{\overline{||}})
=(1,0,0,0)\;\; \text{for $\langle \Lambda|J^s_{\mu,R}|\Lambda_b \rangle$},\,\nonumber\\
&&(e^\uparrow_{||},e^\downarrow_{||},e^\uparrow_{\overline{||}},e^\downarrow_{\overline{||}})
=(0,0,0,0)\;\; \text{for $\langle \Sigma^0|J^s_{\mu,R}|\Lambda_b \rangle$},\,
\end{eqnarray}
we get 
$f_1(0)=g_1(0)=-\sqrt{3/2}\,C_{||}$ for $\langle p|\bar u\gamma_\mu(\gamma_5)b|\Lambda_b \rangle$,
$f_1(0)=g_1(0)=C_{||}$ for $\langle \Lambda|\bar s\gamma_\mu(\gamma_5)b|\Lambda_b \rangle$, and 
$f_1(0)=g_1(0)=0$ for $\langle \Sigma^0|\bar s\gamma_\mu(\gamma_5)b|\Lambda_b \rangle$,
similar to the results based on 
the heavy-quark and large-energy symmetries in Ref.~{\cite{CF}}
for the $\Lambda_b\to (p,\Lambda,\Sigma)$ transitions. When we further extend the study to
the anti-triplet $b$-baryons: $(\Xi_b^-,\Xi_b^0,\Lambda_b^0)$ shown in Table~\ref{f1&g1},
we find that the relation of $f_1=g_1$ is uniquely determined for 
the anti-triplet $b$-baryon transitions.
\begin{table}[t]%[htb]
\caption{Relations between the transition matrix elements.}\label{f1&g1}
\begin{tabular}{|c|c|}
\hline
$\langle {\cal B}_n|(\bar qb)|{\cal B}_b \rangle$ &$f_1(0)=g_1(0)$\\\hline
$\langle p|(\bar ub)|\Lambda_b \rangle$&$-\sqrt{3\over 2}C_{||}$\\
$\langle \Lambda|(\bar ub)|\Xi_b^- \rangle$&${1\over 2}C_{||}$\\
$\langle \Sigma^0|(\bar ub)|\Xi_b^- \rangle$&$-\sqrt{3\over 4} C_{||}$\\
$\langle \Sigma^+|(\bar ub)|\Xi_b^0 \rangle$&$-\sqrt{3\over 2}C_{||}$
\\\hline
%$\langle n|(\bar db)|\Lambda_b \rangle$&---\\
$\langle \Sigma^-|(\bar db)|\Xi_b^- \rangle$&$\sqrt{3\over 2}C_{||}$\\
$\langle \Lambda|(\bar db)|\Xi_b^0 \rangle$&$-{1\over 2}C_{||}$\\
$\langle \Sigma^0|(\bar db)|\Xi_b^0 \rangle$&$\sqrt{3\over 4}C_{||}$
\\\hline
$\langle \Lambda|(\bar sb)|\Lambda_b \rangle$&$C_{||}$\\
$\langle \Sigma^0|(\bar sb)|\Lambda_b \rangle$&0\\
$\langle \Xi^-|(\bar sb)|\Xi_b^- \rangle$&$\sqrt{3\over 2}C_{||}$\\
$\langle \Xi^0|(\bar sb)|\Xi_b^0 \rangle$&$-\sqrt{3\over 2}C_{||}$
\\\hline
\end{tabular}
\end{table}

\section{Numerical Results and Discussions }
For the numerical analysis, 
%the theoretical inputs of %the meson decay constants and
%the Wolfenstein parameters for 
the CKM matrix elements in the Wolfenstein parameterization 
taken from the PDG~\cite{pdg} are given by
\begin{eqnarray}
&&(V_{ub},\,V_{cb})=(A\lambda^3(\rho-i\eta),A\lambda^2)\,,\nonumber\\
&&(V_{cd}=-V_{us},\,V_{cs}=V_{ud})=(-\lambda,1-\lambda^2/2)\,,
\end{eqnarray}
with $(\lambda,\,A,\,\rho,\,\eta)=(0.225,\,0.814,\,0.120\pm 0.022,\,0.362\pm 0.013)$.
The meson decay constants are adopted as
$(f_{\eta_c},\,f_{J/\psi})=(387\pm 7,\,418\pm 9)$ MeV~\cite{Becirevic:2013bsa},
$(f_{D},\,f_{D_s})=(204.6\pm 5.0,\,257.5\pm 4.6)$ MeV~\cite{pdg}, and
$(f_{D^*},\,f_{D^*_s})=(252.2\pm 22.7,\,305.5\pm 27.3)$ MeV~\cite{Lucha:2014xla}.
As given in Ref.~\cite{Hsiao:2014mua}
to explain the branching ratios and CP violating asymmetries of $\Lambda_b\to p(K^-,\pi^-)$, 
we have $|\sqrt{3/2}C_{||}|=0.136\pm 0.009$ 
for $\langle p|\bar u\gamma_\mu (\gamma_5)b|\Lambda_b \rangle$,
which is consistent with the value of  
$0.14\pm 0.03$ in the  light-cone sum rules~\cite{CF} and the theoretical calculations
in Refs.~\cite{Wei:2009np,Gutsche:2013oea}.
With ${\cal B}(\Lambda_b\to J/\psi \Lambda)$ and 
${\cal B}(\Xi_b^-\to J/\psi \Xi^-)$ in Eq.~(\ref{Bb_fBb})
as the experimental inputs, we can estimate the non-factorizable effects
by deviating the color number $N_c=3$ to be between 2 to $\infty$, 
such that we obtain $N_c=2.15\pm 0.17$, 
representing controllable non-factorizable effects ~\cite{Neubert}
with $(a_1,a_2)=(1.00\pm 0.01,0.18\pm 0.05)$ from Eq.~(\ref{a12}).
We list the branching ratios of all possible two-body anti-triplet $b$-baryon decays
in Table~\ref{table1} and Table~\ref{table2}, 
where the uncertainties are fitted with those from
($\rho,\eta,N_c$), the decay constants and $|\sqrt{3/2}C_{||}|$.
%for the hadronic form factors.
{\footnotesize
\begin{table}[b]
\caption{The branching ratios of all possible two-body anti-triplet $b$-baryon decays with $a_1$
%in which the two numbers in each parenthesis are 
fitted by
$N_c=(2.15\pm 0.17,\,\infty$).}\label{table1}
\begin{tabular}{|c|cc|}
\hline
$ M_c = $ & $ D^- $ & $D^{*-}$ \\
\hline
${\cal B}(\Lambda_b \to p M_c)$ 
&  $(6.0 \pm 1.0, 8.2 \pm 1.4) \times 10^{-7}$&  $(1.2 \pm 0.3, 1.6 \pm 0.4) \times 10^{-6}$\\
${\cal B}(\Xi_b^- \to \Lambda M_c)$ 
&  $(1.1 \pm 0.2, 1.5 \pm 0.2) \times 10^{-7}$&  $(2.2 \pm 0.6, 3.0 \pm 0.8) \times 10^{-7}$\\
${\cal B}(\Xi_b^- \to \Sigma^0 M_c)$ 
&  $(3.3 \pm 0.5, 4.5 \pm 0.7) \times 10^{-7}$&  $(6.6 \pm 1.6, 9.0 \pm 2.2) \times 10^{-7}$\\
${\cal B}(\Xi_b^0 \to \Sigma^+ M_c)$ 
&  $(6.3 \pm 1.0, 8.6 \pm 1.4) \times 10^{-7}$&  $(1.3 \pm 0.3, 1.7 \pm 0.4) \times 10^{-6}$\\
\hline\hline
$ M_c = $ & $D_s^-$ & $D_s^{*-}$ \\
\hline
${\cal B}(\Lambda_b \to p M_c)$ 
&  $(1.8 \pm 0.3, 2.5 \pm 0.4) \times 10^{-5}$&  $(3.5 \pm 0.9, 4.7 \pm 1.2) \times 10^{-5}$\\
${\cal B}(\Xi_b^- \to \Lambda M_c)$ 
&  $(3.4 \pm 0.5, 4.6 \pm 0.7) \times 10^{-6}$&  $(6.4 \pm 1.6, 8.8 \pm 2.2) \times 10^{-6}$\\
${\cal B}(\Xi_b^- \to \Sigma^0 M_c)$ 
&  $(9.9 \pm 1.5, 13.6 \pm 2.1) \times 10^{-6}$&  $(1.9 \pm 0.5, 2.6 \pm 0.6) \times 10^{-5}$\\
${\cal B}(\Xi_b^0 \to \Sigma^+ M_c)$ 
&  $(1.9 \pm 0.3, 2.6 \pm 0.4) \times 10^{-5}$&  $(3.6 \pm 0.9, 4.9 \pm 1.2) \times 10^{-5}$\\
\hline
\end{tabular}
\end{table}

\begin{table}[b]
\caption{The branching ratios of all possible two-body anti-triplet $b$-baryon decays with $a_2$
fitted by $N_c=2.15\pm 0.17$.}
\label{table2}
\begin{tabular}{|c|cc|}
\hline
$ M_c = $ & $ D^0 $ & $D^{*0}$ \\
\hline
%${\cal B}(\Lambda_b \to n M_c)$ &  ---&  --- \\
${\cal B}(\Xi_b^- \to \Sigma^- M_c)$ &  $(5.3 \pm 3.3) \times 10^{-5}$&  $(1.1 \pm 0.7) \times 10^{-4}$\\
${\cal B}(\Xi_b^0 \to \Lambda^0 M_c)$ &  $(8.6 \pm 5.3) \times 10^{-6}$&  $(1.7 \pm 1.1) \times 10^{-5}$\\
${\cal B}(\Xi_b^0 \to \Sigma^0 M_c)$ &  $(2.5 \pm 1.6) \times 10^{-5}$&  $(5.0 \pm 3.4) \times 10^{-5}$\\
${\cal B}(\Lambda_b \to \Lambda M_c)$ &  $(1.6 \pm 1.0) \times 10^{-6}$&  $(3.3 \pm 2.2) \times 10^{-6}$\\
${\cal B}(\Lambda_b \to \Sigma^0 M_c)$ &  $0$&  $0$ \\
${\cal B}(\Xi_b^- \to \Xi^- M_c)$ &  $(2.7 \pm 1.7) \times 10^{-6}$&  $(5.5 \pm 3.6) \times 10^{-6}$\\
${\cal B}(\Xi_b^0 \to \Xi^0 M_c)$ &  $(2.6 \pm 1.6) \times 10^{-6}$&  $(5.2 \pm 3.5) \times 10^{-6}$\\
\hline
\end{tabular}
%\end{table}

%\begin{table}[b]
%\caption{($a_2=0.178256$, $a_2=-0.365$.)}
%\label{fLb_R}
%\renewcommand\arraystretch{2}
\begin{tabular}{|c|cc|}
\hline
$ M_c = $ & $\eta_c$ & $J/\psi$ \\
\hline
%${\cal B}(\Lambda_b \to n M_c)$ &  ---&  --- \\
${\cal B}(\Xi_b^- \to \Sigma^- M_c)$ &  $(1.4 \pm 0.8) \times 10^{-5}$&  $(2.9 \pm 1.8) \times 10^{-5}$\\
${\cal B}(\Xi_b^0 \to \Lambda^0 M_c)$ &  $(2.3 \pm 1.4) \times 10^{-6}$&  $(4.7 \pm 2.9) \times 10^{-6}$\\
${\cal B}(\Xi_b^0 \to \Sigma^0 M_c)$ &  $(6.6 \pm 4.1) \times 10^{-6}$&  $(1.4 \pm 0.8) \times 10^{-5}$\\
${\cal B}(\Lambda_b \to \Lambda M_c)$ &  $(1.5 \pm 0.9) \times 10^{-4}$&  $(3.3 \pm 2.0) \times 10^{-4}$\\
${\cal B}(\Lambda_b \to \Sigma^0 M_c)$ &  $0$ &  $0$ \\
${\cal B}(\Xi_b^- \to \Xi^- M_c)$ &  $(2.4 \pm 1.5) \times 10^{-4}$&  $(5.1 \pm 3.2) \times 10^{-4}$\\
${\cal B}(\Xi_b^0 \to \Xi^0 M_c)$ &  $(2.3 \pm 1.4) \times 10^{-4}$&  $(4.9 \pm 3.0) \times 10^{-4}$\\
\hline
\end{tabular}
\end{table}
}
The decay branching ratios in Table~\ref{table1}
%of the decay modes with 
are given by $a_1$ with $N_c=(2.15\pm 0.17,\,\infty)$ as the theoretical inputs
%where the two sets of values for $a_1$ with $N_c=2.15\pm 0.17$ and $N_c=\infty$,
%respectively, 
to demonstrate the insensitive non-factorizable effects. Note that
$N_c=2.15\pm 0.17$ is fitted from ${\cal B}(\Lambda_b\to J/\psi \Lambda)$ and 
${\cal B}(\Xi_b^-\to J/\psi \Xi^-)$, while $N_c=\infty$ results in $a_1\simeq c_1^{eff}$, wildly used in the generalized factorization.
As the first measurement for the color-allowed decay mode,
the predicted ${\cal B}(\Lambda_b\to D_s^- p)=
(1.8\pm 0.3)\times 10^{-5}$ or $(2.5 \pm 0.4)\times 10^{-5}$ in Table~\ref{table1}
seems to disagree with the data in Eq.~(\ref{Bb_fBb}). Nonetheless, 
the predicted numbers driven by $a_1$ can be reliable
as it is insensitive to the non-factorizble effects,
whereas the data with the upper bound has a large uncertainty.
Despite the color-allowed modes, 
the decay branching ratios of $D^{(*)-}$ are found
to be 30 times smaller than the $D^{(*)-}_s$ counterparts. This can be simply understood 
by the relation of $(V_{cd}/V_{cs})^2(f_{D^{(*)}}/f_{D_s^{(*)}})^2\simeq 0.03$.
It is also interesting to note that the vector meson modes are 2 times
as large as their pseudoscalar meson counterparts.

For the decay modes driven by $a_2$ as shown in Table~\ref{table2}, we only list the results
with $a_2=0.18\pm 0.05$ ($N_c=2.15\pm0.17$). The reason is that 
$a_2\simeq c_2^{eff}= -0.365$ with $N_c=\infty$ yields
${\cal B}(\Lambda_b\to J/\psi \Lambda)=(1.4\pm 0.2)\times 10^{-3}$ and 
${\cal B}(\Xi_b^-\to J/\psi \Xi^-)=(2.1\pm 0.3)\times 10^{-3}$,
which are in disagreement with the data in Eq.~(\ref{Bb_fBb}),
demonstrating that the decays are sensitive to the non-factorizable effects. 
From Table~\ref{table2}, we see that
both ${\cal B}(\Lambda_b\to J/\psi \Lambda)$ and ${\cal B}(\Xi_b^-\to J/\psi \Xi^-)$
are reproduced to agree with the data in Eq.~(\ref{Bb_fBb}) within errors.
Note that 
${\cal B}(\Lambda_b\to J/\psi \Lambda)/{\cal B}(\Xi_b^-\to J/\psi \Xi^-)
\simeq 0.65$ in our calculation results from $(C_{||})^2/(\sqrt {3/2}C_{||})^2\simeq 0.67$
as the ratio of their form factors in Table~\ref{f1&g1},
which is in accordance with the $SU(3)$ flavor and $SU(2)$ spin symmetries.
The more precise measurement of this ratio in the future will test the validity  of the symmetries.
As ${\cal B}(\Xi_b^-\to J/\psi \Xi^-)=O(10^{-4})$, we emphasize that 
more experimental searches should be done for the two-body $\Xi_b$ decays,
while most of the recent observations are from the $\Lambda_b$ decays. 
Since ${\cal B}(\Lambda_b \to \Sigma^0 M_c)=0$ results from 
the $SU(3)$ flavor and $SU(2)$ spin symmetries as well as the heavy flavor symmetry,
its non-zero measurement will imply for the broken symmetry.
Through the $b\to c\bar cs$ transition at the quark level, 
${\cal B}(\Lambda_b \to \Lambda M_c)$, $\mathcal{B}(\Xi_b^- \to \Xi^- M_c)$ and 
${\cal B}(\Xi_b^0 \to \Xi^0 M_c)$ with $M_c=\eta_c$ and $J/\psi$ are all $O(10^{-4})$
as shown in the right-bottom of Table~\ref{table1}. 
In contrast, the neutral $D^{(*)0}$ modes via the $b\to c\bar c d$ transition
have the branching ratios of order $10^{-6}$ caused by
the suppression of $(V_{cb}V_{cd})^2/(V_{cb}V_{cs})^2\simeq 0.225^2$.
Finally, we remark that 
${\cal B}(\Xi_b^- \to \Xi^- M_c)\simeq \mathcal{B}(\Xi_b^0 \to \Xi^0 M_c)$
is due to the isospin symmetry. 

\section{Conclusions}
In sum, we have studied all possible anti-triplet ${\cal B}_b$ decays of 
the two-body charmful ${\cal B}_b\to {\cal B}_n M_c$ decays.
We have found ${\cal B}(\Lambda_b\to D_s^- p)=(1.8\pm 0.3)\times 10^{-5}$,
which is within the measured upper bound of 
${\cal B}(\Lambda_b\to D_s^- p)<4.8(5.3)\times 10^{-4}$ at 90\%\,(95\%) C.L.,
and reproduced
${\cal B}(\Lambda_b\to J/\psi \Lambda)=(3.3\pm 2.0)\times 10^{-4}$ and 
${\cal B}(\Xi_b^-\to J/\psi \Xi^-)=(5.1\pm 3.2)\times 10^{-4}$
in agreement with the data. Moreover, we have predicted 
${\cal B}(\Lambda_b \to \Lambda \eta_c)=(1.5 \pm 0.9) \times 10^{-4}$,
${\cal B}(\Xi_b^- \to \Xi^- \eta_c)=(2.4 \pm 1.5) \times 10^{-4}$, and 
${\cal B}(\Xi_b^0 \to \Xi^0 \eta_c,\Xi^0 J/\psi)=(2.3 \pm 1.4,4.9 \pm 3.0) \times 10^{-4}$,
which are accessible to the experiments at the LHCb, while
${\cal B}(\Xi_b^- \to \Xi^- M_c)\simeq {\cal B}(\Xi_b^0 \to \Xi^0 M_c)$ is due to
the isospin symmetry.

%\newpage
\section*{ACKNOWLEDGMENTS}
The work was supported in part by National Center for Theoretical Science, National Science
Council (NSC-101-2112-M-007-006-MY3 and NSC-102-2112-M-471-001-MY3), 
MoST (MoST-104-2112-M-007-003-MY3) and National Tsing Hua
University (104N2724E1).

\end{document}